# Utilisation d'accéléromètres pour évaluer l'activité physique des truies gestantes logées en groupes
## Développement de la méthode et utilisation dans six élevages au DAC

**Yannick Ramonet, Carole Bertin**
Chambre d'agriculture de Bretagne. Avril 2015

*Les truies en groupes, logées au DAC, sont debout en moyenne 4h19min par jour, et changent 29 fois de posture. Une méthode a été développée pour mesurer l'activité des truies à l'aide de capteurs placés sur leur patte. Utilisée dans six élevages de production, les résultats illustrent la diversité du niveau d'activité entre les élevages et entre les animaux au sein des groupes. Ce paramètre est essentiel dans une perspective d'alimentation précise des truies en tenant compte du besoin énergétique lié à cette activité.*

## 1. Introduction

Connaître l'activité motrice de chaque truie doit permettre d'améliorer la connaissance du besoin alimentaire de l'animal. En effet, le besoin énergétique d'entretien est doublé lorsque la truie est debout par rapport à une position couchée (Noblet et al., 1994). Des changements du niveau d'activité peuvent également refléter un problème de santé.

Les données disponibles sur l'activité des truies gestantes logées en groupes sont peu nombreuses. Des observations en élevages nous ont permis d'évaluer les distances parcourues par des truies en groupes sur une durée de 6 heures (Ramonet et Tertre, 2014). L'activité des animaux apparait dépendante du mode de logement. Les truies logées au DAC dynamique, dans des salles de 400 à 500 m² parcouraient en moyenne 362 m au cours des 6 heures d'observation. Logées en petites cases avec bat-flanc, la distance parcourue n'est plus que de 50 m en moyenne. Les logements en réfectoire-courette et au DAC stable occupent une position intermédiaire. Nos travaux montraient également une importante disparité entre les truies au sein d'un même groupe, les différences étant les plus importantes pour les animaux en DAC dynamique.

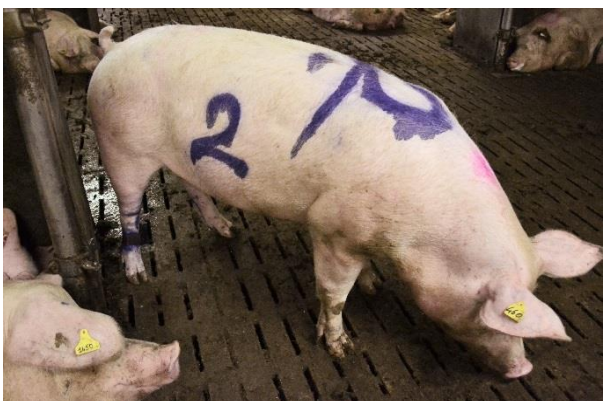

Les éleveurs ont peu de notions sur le niveau d'activité réel de leurs truies. Ceci est d'autant plus difficile que les animaux sont nombreux dans les élevages, et qu'une part importante de l'activité se déroule en l'absence de l'éleveur dans le bâtiment. C'est par exemple le cas pour le DAC lorsque le cycle alimentaire débute en fin d'après-midi ou au cours de la nuit. Le déplacement des animaux qui vont s'alimenter a alors lieu en l'absence de présence humaine dans l'élevage. Les présentations de notre précédent travail (Ramonet et Tertre, 2014) où nous affichons des parcours des truies dans les salles de gestation, complétées par la réalisation de vidéos, suscitent toujours des discussions, voire une certaine surprise de la part des éleveurs.

Evaluer l'activité physique des animaux est une tâche complexe. Les observations du comportement des truies, qu'il s'agisse d'observations directes dans le bâtiment ou de dépouillement d'enregistrements vidéos, sont nécessairement limitées dans le temps. La dimension importante des cases pour les truies gestantes, notamment en présence d'un DAC complique également le suivi des animaux. Quatre caméras sont par exemple nécessaires pour suivre le mouvement des truies dans des salles de 170 à 490 m² (Jensen et al., 2000).

L'utilisation de capteurs placés sur l'animal est une voie intéressante pour obtenir des données sur de longues périodes. Le principe de la pose de capteurs sur les animaux est utilisée sur le veau, la vache ou le cheval (Passillé et al., 2010 ; Mattachini et al., 2011 ; Burla et al., 2014) et des capteurs à placer autour du cou de la vache laitière sont déjà commercialisés. De manière plus anecdotique, la pose de balise sur des animaux sauvages est une pratique courante pour suivre les déplacements de cigognes, de caribous ou autres méduses. En revanche, peu de références sont disponibles chez le porc, et le développement se fait pour le moment essentiellement pour des projets de recherche. Des travaux portent également sur la localisation de l'animal au sein de grandes salles. Cette technique repose sur le repérage par triangulation du signal renvoyé par une puce placée sur l'animal. Cette technique est employée pour repérer notamment les vaches dans de grandes étables, mais a également été testée chez la truie.

Des références sont disponibles chez la truie à l'aide d'accéléromètres placés essentiellement au cou (Cornou et Lundbye-Christensen, 2008) ou sur la patte de la truie logée en stalle individuelle (Ringgenberg et al., 2010). Ces mesures ont été réalisées sur un nombre réduit d'animaux dans le but de travailler la méthode de détection des comportements de l'animal.

L'objectif de la présente étude consiste à mesurer, à l'aide d'accéléromètres, le temps passé debout et couché de truies logées en groupes. Une première phase du travail consiste à mettre au point une méthode de fixation des capteurs et de standardiser l'analyse du signal. Dans un second temps, les capteurs sont placés sur 12 à 13 truies dans six élevages de production équipés de DAC.

# 1. Mise au point et de validation de la méthode de mesure

## 1.1. Accéléromètres

Les accéléromètres utilisés sont des HOBO® Pendant G data logger (Figure 1). Il s'agit d'un enregistreur de gravité, accélération, vibration et déplacements angulaires sur 1, 2 ou 3 axes. L'échelle d'accélération est de ± 3 g.

Des boîtiers métalliques en acier et en aluminium (dimension : 80×35×30 mm ; poids : 145 ± 57 g) ont été spécialement conçus pour contenir et protéger les accéléromètres.

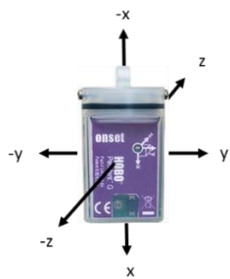

**Figure 1 : accéléromètres utilisés dans le cadre de l'étude**

## 1.2. Fixation sur l'animal

La mise au point du mode de fixation des boîtiers contenant les accéléromètres sur l'animal a été réalisée sur les truies logées au DAC à la station porcine de Guernevez.

Le système de fixation devait répondre à 4 objectifs posés au départ du projet :
- Pouvoir être fixé sur des truies logées en groupes et libres de se déplacer
- Etre suffisamment robuste pour rester en place pour une durée de 3 à 4 jours au moins
- Ne pas blesser l'animal
- Etre suffisamment stable pour que le signal enregistré par l'accéléromètre soit net pour faciliter son interprétation.

Nous avons privilégié dès le départ la fixation du capteur à la patte de la truie. Le membre de l'animal est à la vertical lorsque l'animal est debout et à l'horizontal lorsqu'il est couché. Cette fixation devait nous permettre d'obtenir un signal net issu de l'accéléromètre.

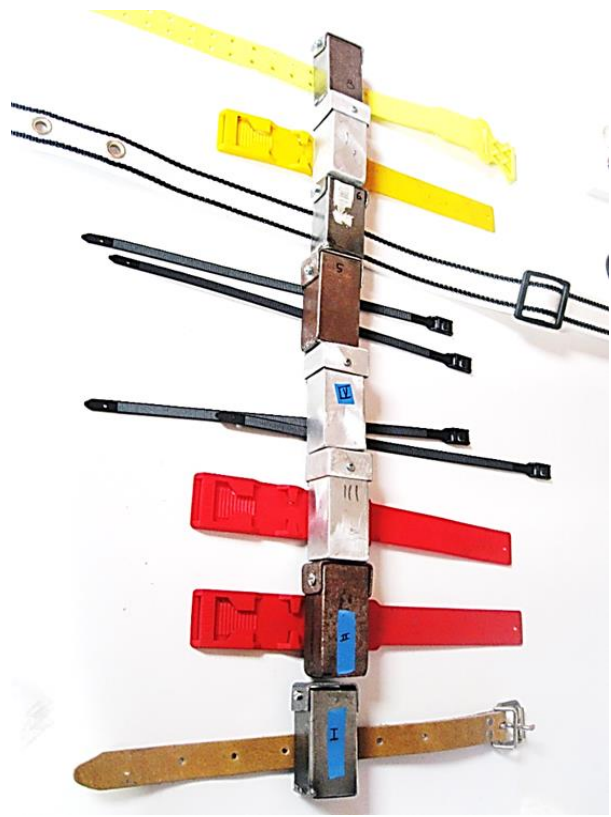

**Figure 2 : Types de fixation testés à la patte des truies**

Aucun système de fixation conçu pour être fixé à la patte des truies n'étant disponible, nous avons adapté des équipements développés pour d'autres animaux, ainsi que du matériel de quincaillerie.

Le matériel testé (Figure 2), fixé la patte arrière de la truie, était constitué de :

*Accéléromètres pour la mesure de l'activité physique des truies en groupes*

- Bracelets polyuréthane pour vaches raccourcis à la dimension d'une patte de truie
- Collier ovin
- Collier cuir canin
- Collier de serrage (type Colson)

Un collier bovin a également été posé autour du cou d'une seule truie. Ces différents systèmes ont été fixés sur 8 truies.

Le boîtier et la patte de l'animal sont couverts d'un produit utilisé pour l'hygiène cutanée des porcs (Cicalm®) possédant des propriétés amérisantes répulsives. L'objectif consistait à limiter le léchage ou la morsure du boîtier par les autres truies du groupe au cours des premières heures suivant la fixation.

Le mode de fixation s'est révélé satisfaisant. Des capteurs ont été perdus au bout de 5 et 6 jours de fixation, deux fixés à l'aide de bracelets polyuréthane, et un fixé avec un collier de serrage. En revanche, la peau de la patte de certaines truies était légèrement irritée suite au frottement avec le boîtier métallique.

A la suite de cet essai, le dispositif a été adapté en protégeant la patte de la truie avec une mousse de 13 mm d'épaisseur.

Le mode de fixation retenu est constitué de 2 colliers de serrage en plastique de 9 mm de largeur (Figures 3 et 4), serrés à l'aide d'une pince spécifique. La patte et le boîtier sont couverts de Cicalm®.

Les colliers de serrage présentaient plusieurs avantages par rapport à notre projet : serrage facile à l'aide de la pince, possibilité d'adapter la pression de serrage, coût faible. Inversement, plusieurs raisons expliquent la non rétention des autres modes de fixation à la patte.

- Bracelets polyuréthane : nécessité de le raccourcir pour l'adapter à la dimension de la patte. Dispositif à usage unique, mâchonné par les truies. Coût élevé dans le cas d'un déploiement important.
- Collier ovin : fixation solide, mais compliqué sur des truies en mouvement
- Collier cuir canin : serrage et retrait difficile.

Un collier bovin posé autour du cou d'une truie est resté en place pendant les 6 jours de l'essai. En revanche le signal issu de l'accéléromètre était compliqué à interpréter comparativement à celui obtenu au niveau de la patte. Pour cette raison, nous avons choisi de ne pas le conserver pour la suite de l'étude.

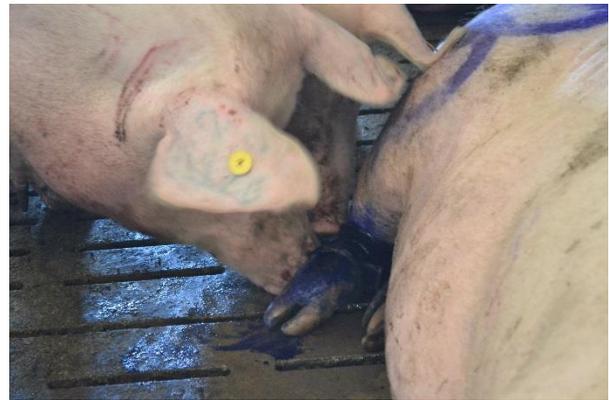

**Figure 4 : Le capteur à la patte d'une truie équipée intéresse les autres animaux du groupe, ce qui rend la protection nécessaire**

### 1.3. Validation du dispositif de mesure

#### 1.3.1. Animaux, logement

Douze truies logées sur caillebotis intégral avec alimentation par DAC à la station de Guernevez ont été retenues pour la phase de validation. Les animaux ne présentaient pas de troubles locomoteurs. Les boîtiers contenant les capteurs étaient fixés sur ces truies selon le protocole précédemment décrit. La dimension de la case est de 70 m² pour une capacité de 23 truies conduites dans un groupe dynamique. La séquence alimentaire débutait à minuit.

#### 1.3.2. Observations comportementales

Afin de valider les enregistrements des accéléromètres, l'observation du comportement des truies a été utilisée en référence. Les animaux ont été observés pendant trois jours consécutifs pendant une durée de trois heures, en début de matinée. L'heure de chaque changement de posture (debout/couché) était notée. Au total,

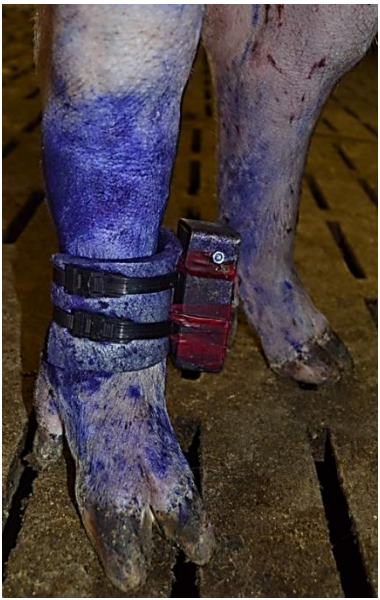

**Figure 3 : Boîtier métallique contenant les accéléromètres, installé sur la patte arrière d'une truie, couvert de Cicalm®**

L'équipement est fixé sur les truies logées en groupes, libres de se déplacer dans la case. Une poignée d'aliment pouvait être donnée à la truie pour limiter ses déplacements le temps de la fixation. L'installation du matériel durait de 20 à 40 secondes environ. Le retrait des boîtiers était réalisé en sectionnant les deux colliers de serrage à l'aide d'un sécateur.

nous disposons donc de 3 heures × 12 truies × 3 jours d'observations du comportement utilisées comme données de référence. Ces observations étaient réalisées le matin, entre 8h et 11h, à un moment où les truies étaient actives. Un enregistrement video à l'aide de 2 caméras complétait le dispositif et était utilisé pour vérifier les observations directes en cas de doute sur l'heure du changement de posture.

### 1.3.3. Programmation des capteurs

Le choix du nombre d'axes et la fréquence d'enregistrement des données par l'accéléromètre était une question essentielle pour ce projet. L'accéléromètre possède en effet une capacité de mémoire limitée. Plus on enregistrait de données, moins la durée totale d'enregistrement était élevée.

A l'issu d'un premier test, où des enregistrements étaient réalisés sur les 3 axes X, Y et Z, nous avons constaté que l'axe vertical X était suffisant pour discriminer les postures debout/couché. Au final, seul cet axe vertical est retenu.

Plusieurs fréquences d'enregistrement ont été testées, à 10, 20 et 30 secondes. A ces intervalles la valeur de l'axe X, exprimée en g, est enregistrée. L'intervalle de 30 secondes apparaît trop long. L'analyse des données a montré qu'avec un enregistrement toutes les 20 secondes la qualité de la mesure était équivalente à un enregistrement toutes les 10 secondes. Avec cette fréquence de mesure, tous les changements de posture devraient être enregistrés.

Les paramètres retenus pour la programmation de l'accéléromètre sont
- un enregistrement des données toutes les 20 secondes
- sur le seul axe X vertical.

Cette programmation permet de discriminer de manière satisfaisante les positions debout et couchée. Avec ces paramètres, la mémoire interne de l'accéléromètre permettrait d'enregistrer les données sur une durée totale de 2 semaines. Pour la phase de validation, les boîtiers ont été récupérés à l'issue de 3 jours d'enregistrement, et les données transférées sur un tableur Excel.

### 1.3.4. Traitement du signal

Le fichier de données brutes obtenu sur Excel est constitué de la valeur de l'axe X toutes les 20 secondes, exprimée en g. Le capteur est en position verticale lorsque la valeur X est égale à 1 g, et en position horizontale lorsque la valeur X est égale à 0 g.

En situation de mesure, la valeur de l'axe X peut varier entre -3 g et + 3 g selon le mouvement appliqué au capteur. Il convient donc d'appliquer un traitement du signal dans le but d'obtenir les seules références debout-couché en enlevant les artéfacts.

Deux règles principales ont été retenues pour analyser le signal issu du capteur :

- Le seuil retenu au-dessus duquel une truie est considérée debout est fixé à 0,65 g. Cette valeur est celle qui offre les meilleurs résultats en termes de discrimination des changements de posture. Nous avons testé plusieurs valeurs (0,55-0,65-0,75-0,85). Avec un seuil de 0,65 g, seulement 2 changements de position en moyenne n'ont pas été détectés sur une durée de 3 heures, pour une durée moyenne de la séquence de 1 min 50 sec.

- un animal est considéré debout ou couché si cette période comporte au moins deux points consécutifs dans une même posture. Cette règle permet d'éliminer des artéfacts du signal et correspond à une réalité comportementale dans la mesure où nous n'avons pas observé de truie qui se levait et se recouchait en une vingtaine de secondes. Un algorithme avec ces deux règles de décision a été développé sur Excel et permet d'obtenir le traitement illustré par la Figure 5.

La qualité de la mesure est exprimée par la sensibilité (SE) et la spécificité (SP). La sensibilité SE mesure la capacité de la méthode à donner un résultat positif lorsque l'hypothèse est vérifiée. Autrement dit, elle mesure la probabilité qu'une truie changeant de posture soit détectée par l'accéléromètre. La spécificité SP mesure la capacité d'une méthode à donner un résultat négatif lorsque l'hypothèse n'est pas vérifiée. Elle mesure donc la probabilité qu'une truie qui ne change pas de posture ne soit pas détectée par l'accéléromètre.

**Tableau 1 : Sensibilité (SE) et spécificité (SP) des mesures de posture pour les 12 truies au cours d'un suivi de 9 heures**

|  |  | Méthode de référence : Observation directe | |
|---|---|---|---|
|  |  | [+] | [-] |
| Méthode testée : Accéléromètre | [+] | Vp = 176 | Fp = 28 |
|  | [-] | Fn = 2 | Vn = 19234 |
|  |  | Sensibilité 98,8 % | Spécificité 99,8 % |

*Test réalisé pour 19440 points de mesures où la truie change de posture [+] ou ne change pas de posture [-]. Vp : Vrai Positif ; Fp : Faux positif ; Fn : Faux négatif ; Vn : Vrai négatif.*

Ces valeurs sont calculées au cours des 9 heures d'observation en utilisant les observations directes comme référence pour chacun des 19440 points de mesures réalisés par les accéléromètres. Chacun de ces points correspond à un enregistrement réalisé toutes les 20 secondes sur les 12 truies. Pour chaque mesure, la donnée est classée comme vrai ou faux positif (Vp, Fp) ou vrai ou faux négatif (Vn, Fn). La sensibilité SE mesure le taux de vrais positifs (SE = Vp/(Vp+Fn)) et la spécificité SP le taux de vrais négatifs (SP=Vn/(Vn+Fp)).

Au cours des 9 heures d'observation, 178 changements de posture ont été observés pour les 12 truies pour une moyenne de 14,8 changements par truie (mini = 9 ; maxi = 25). Seulement deux changements de posture n'ont pas été détectés par l'accéléromètre. La sensibilité de la mesure s'élève à 98,8 ± 2,7% (Tableau 1), en étant de 100% pour 11 des 12 truies. La spécificité de la mesure est de 99,8 ± 0,1%, le capteur ayant mesuré au total 28 changements répartis sur 10 truies, et que l'observateur n'avait pas notés

**Figure 5 : Exemple d'enregistrement et de traitement du signal lors de la phase de validation**

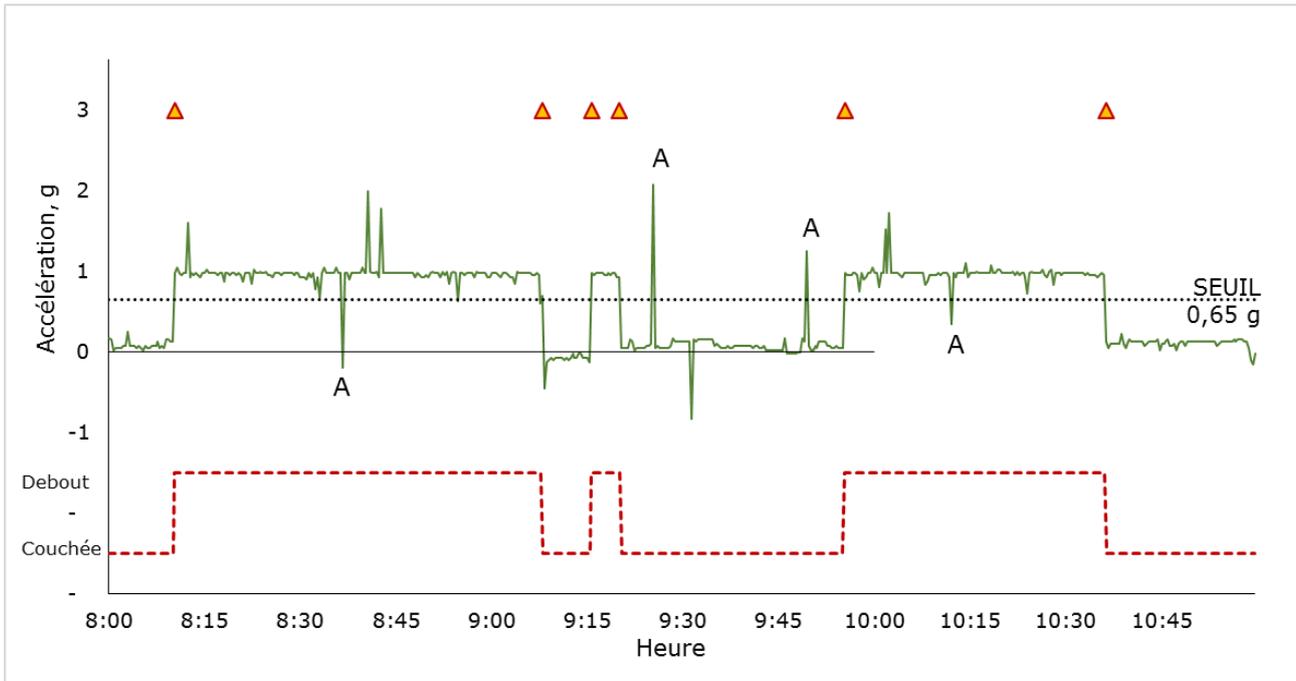

— *donnée brute de l'accélération enregistrée par l'accéléromètre toutes les 20 s. sur une échelle ± 3 g. Traitement du signal à partir d'un seuil à 0,65 g et l'élimination des artéfacts (A).* --- *Profil des postures debout/couchée au cours de la période. L'observation des changements de posture (▲) est utilisée comme référence.*

## 2. Mesure de l'activité des truies en élevage

### 2.1. Elevages, animaux et capteurs

Des mesures ont été réalisées dans six élevages de production, identifiés A à F, équipés de DAC et avec un sol en caillebotis, (Tableau 2, Figure 6). Ils se distinguent par le type et la taille des groupes (3 groupes stables et 3 groupes dynamiques), la dimension des cases, les équipements pour l'alimentation (marques commerciales des DAC) et l'aménagement intérieur.

Les boîtiers contenant les accéléromètres ont été posés en début de semaine sur des truies maintenues en groupes pour une durée de 2 à 4 jours. Chaque matin, un intervenant passait vérifier le bon déroulement de l'étude.

Treize truies ont été équipées dans les élevages A et B, 12 dans les quatre autres. Dans chaque élevage, les animaux retenus appartenaient à la même bande de truies. Elles étaient en milieu de gestation (7 à 10 semaines après insémination), ne présentaient pas de trouble locomoteur, et ont été choisies pour représenter une diversité de rangs de portée. Dans les élevages C et D, les cochettes étaient logées dans un autre groupe et ne sont donc pas présentes dans l'étude.

### 2.2. Traitement des données issues des accéléromètres

Les données issues des accéléromètres ont été traitées sur Excel en utilisant l'algorithme précédemment décrit.

Les données calculées sont exprimées par séquences de 24 heures. Le rythme d'activité de la truie est en partie lié à son alimentation, et l'heure de démarrage de la séquence alimentaire était très différente entre les élevages (Tableau 2). Pour obtenir une base équivalente entre les élevages, la notion de « jour » a été établie pour un début de journée qui démarrait 1 heure avant le début de la séquence alimentaire, alors que l'activité des truies est généralement faible.

*Accéléromètres pour la mesure de l'activité physique des truies en groupes*

Les variables calculées pour chacun des truies sont : le temps passé debout (TpsD) et couché (TpsC) par jour, le nombre de changement de posture par jour (CPost), la durée moyenne des séquences debout (SeqMoyD) et couchée (SeqMoyC), la durée maximale d'une séquence debout (SeqMaxD) et couchée (SeqMaxC).

**Tableau 2 : Description des élevages retenus pour les mesures d'activités par accéléromètre**

|  | Elevages | | | | | |
|---|---|---|---|---|---|---|
|  | **A** | **B** | **C** | **D** | **E** | **F** |
| Type de conduite du groupe | Stable | Stable | Stable | Dynamique | Dynamique | Dynamique |
| Surface de la case (m²) | 72 | 138 | 84 | 282 | 285 | 707 |
| Nombre de truies du groupe | 30 | 60 | 36 | 130 | 115 | 330 |
| Heure de démarrage de la séquence alimentaire | 5h00 | 15h30 | 16h30 | 17h30 | 12h00 | 15h00 |
| Température moyenne (°C) | 21,0 ± 1,4 | 23,4 ± 0,8 | 23,0 ± 0,7 | 22,3 ± 0,6 | 23,9 ± 0,8 | 22,9 ± 1,9 |
| Rang de portée moyen [min-max] | 2,9 [0 – 6] | 4,1 [0 – 7] | 4,2 [1 – 8] | 3,5 [1 – 7] | 3,9 [0 – 11] | 2,5 [0 – 7] |
| Nombre de jours de mesures | 2 | 4 | 4 | 3 | 4 | 3 |
| Nombre d'accéléromètres |  |  |  |  |  |  |
| installés | 13 | 13 | 12 | 12 | 12 | 12 |
| en place à J1 | 10 | 10 | 12 | 11 | 11 | 11 |
| en place à J2 | 10 | 9 | 12 | 11 | 11 | 11 |
| en place à J3 | - | 8 | 12 | 11 | 11 | 10 |
| en place à J4 | - | 7 | 12 | - | 11 | - |

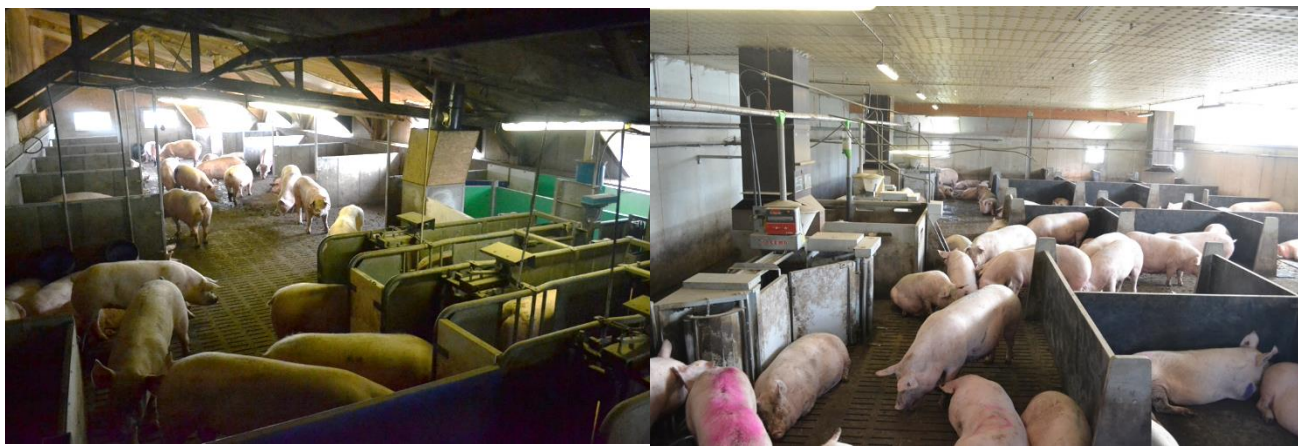

**Figure 6 : Deux des élevages retenus pour les mesures. B : Elevage avec DAC stable, 60 truies dans le groupe / D : DAC dynamique, 130 truies dans le groupe.**

### 2.3. Analyses statistiques

Les données sont traitées à l'aide du logiciel R (2014). On teste grâce à un modèle mixte les effets du jour, de l'élevage, du rang, de la truie (effet aléatoire hiérarchisé dans l'effet élevage), et les interactions jour-élevage, rang-jour et rang-élevage sur TpsD et Cpost.

Une classification a permis de caractériser les truies du point de vue de l'ensemble des mesures (analyse en composante principale suivie d'une classification ascendante hiérarchique CAH, package FactoMineR). Un test du $\chi^2$ d'indépendance a permis de déterminer si les classes de truies mises en évidence par la CAH étaient représentées dans tous les élevages ou étaient associées à une conduite spécifique (fonction catdes de FactoMineR).

## *2.4. Résultats*

### 2.4.1. Tenue des capteurs

74 truies ont été équipées d'accéléromètres entre les 6 élevages. Le lendemain de la fixation, 6 boîtiers ont été retrouvés au sol et 3 volontairement retirés dans le premier élevage équipé. Ils ont soit glissé sur la patte de la truie soit ils ont été arrachés par mordillement. Par la suite, peu de capteurs sont tombés et la tenue était de meilleure qualité dans les 4 derniers élevages équipés : C (aucune perte sur 4 jours), D et E (1 perte à J1) et F (2 pertes à J1 et à J3). Notre technicité de pose et de serrage des colliers s'est améliorée au cours du temps.

Des données sur une durée minimale de 24h sont obtenues sur 65 truies. L'activité des truies a été mesurée à l'aide des accéléromètres sur une durée totale de 5064 heures.

### 2.4.2. Temps passé debout et changements de posture

Sur l'ensemble des animaux, le temps moyen passé debout par jour est de 4h19 ± 1h54, avec une importante variation interindividuelle comprise entre 1h16 et 9h22 par truie. Le nombre moyen de changements de posture par séquence de 24 heures (CPost) est de 29 ± 12 et varie entre 8 et 76 selon les individus.

Les interactions jour-élevage et jour-rang et l'effet jour sont non significatifs sur ces deux variables (p>0,05).

En revanche, les effets de l'élevage, du rang et de la truie sur TpsD et Cpost sont significatifs, ainsi que l'interaction entre le rang et l'élevage (p<0,05).

**Figure 7 : Dispersion du temps passé debout par les truies des 6 élevages (A à F)**

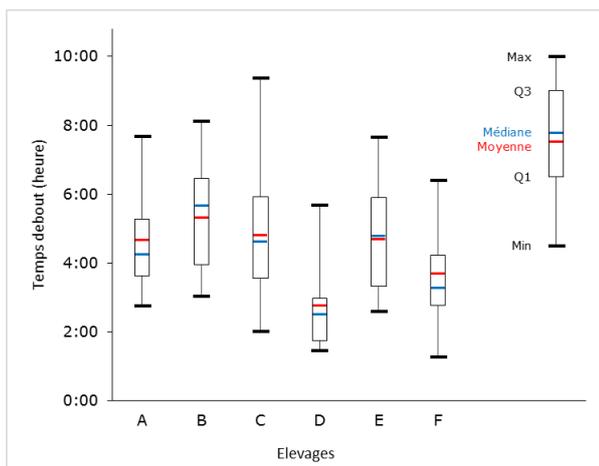

La figure 7 illustre la dispersion sur le temps passé debout entre les truies dans les six élevages. Les différences sont importantes entre les élevages et entre les truies au sein d'un même élevage. Par exemple dans l'élevage D les truies sont debout en moyenne 2h45 par jour contre 5h18 dans l'élevage B. Dans l'élevage C, la truie la plus active est debout 9h22 par jour en moyenne contre 2h01 pour la truie la moins active.

### 2.4.1. Variabilité individuelle et répétabilité journalière

L'interaction truie-jour ne peut pas être testée car nous n'avons qu'une mesure journalière par truie. Pour évaluer la répétabilité de l'activité des truies, on représente dans la figure 8 les mesures de TpsD par truie, élevage et jour. Elle permet de mettre en évidence l'effet de l'élevage (p=0,002), la grande diversité des truies (effet truie significatif, p<0,001) et montre que l'activité de la majorité des truies est répétable d'une journée à l'autre. La variabilité des mesures due à l'effet truie est bien plus importante que la variabilité due au jour. Ce résultat est similaire sur la variable CPost.

### 2.4.2. Effet du rang de portée

L'effet du rang de portée sur TpsD et CPost change en fonction de l'élevage. La figure 9 montre de l'effet du rang sur TpsD pour les élevages B et E, correspondant aux deux scenarii observés. Le rang n'a pas d'effet sur TpsD dans les élevages A, B, D et F. En revanche, dans les élevages C et E, le temps debout augmente avec le rang de portée (p<0,05).

Pour le nombre de changements de posture, lorsque l'effet rang existe, CPost diminue lorsque le rang augmente.

**Figure 9 : Temps passé debout (TpsD, minutes/24 heures) en fonction du rang de portée pour les truies des élevages B et E.**

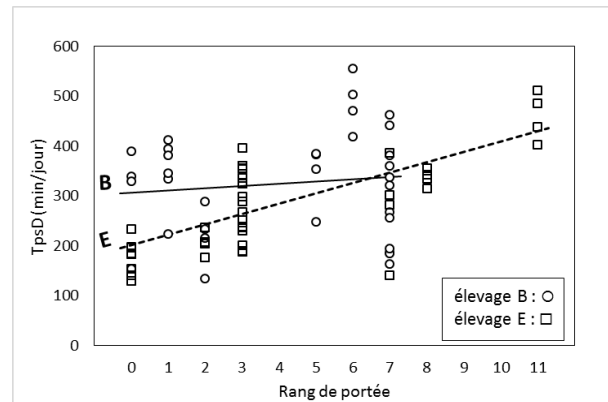

**Figure 8 – Evolution du temps passé debout (TpsD, minutes/24 heures, en ordonnée) pour chacune des truies dans les six élevages A à F, au cours de 2 à 4 jours d'enregistrement (J1 à J4 en abscisse).**

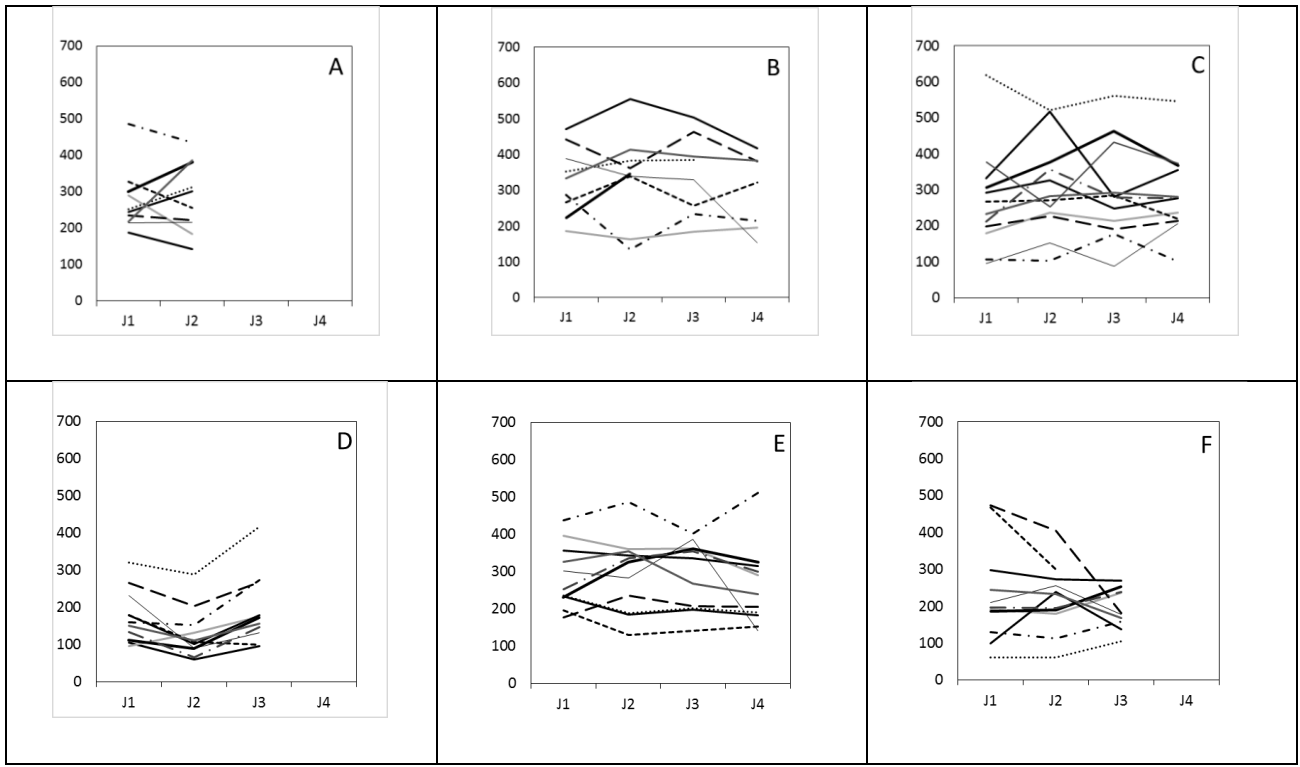

**Tableau 3 – Caractéristiques des cinq classes de truies (valeurs moyennes)[1]**

|  | Moyenne de l'ensemble | Classes | | | | |
|---|---|---|---|---|---|---|
|  |  | C1 | C2 | C3 | C4 | C5 |
| Nombre de truies | 0 | 14 | 21 | 21 | 6 | 3 |
| TpsD | 04h19 | 03h53 | **02h52*** | **05h26*** | 04h22 | **07h40*** |
| SeqMoyD | 00h22 | **00h11** | **00h14** | 00h25 | 00h32 | **01h16*** |
| SeqMoyC | 01h35 | **00h58*** | 01h45 | 01h23 | **02h27*** | **02h25*** |
| SeqMaxD | 01h42 | **00h55*** | **00h56** | **02h20** | 02h00 | **05h21*** |
| SeqMaxC | 06h35 | **05h22** | 06h55 | 05h48 | **10h08*** | 06h43 |
| CPost | 29 | **43*** | 26 | 29 | **18** | **14*** |
| Rang Portée | 3.5 | **2.2*** | 3.2 | 3.7 | **6.5** | 5 |
| Nombre de truies par élevage A/B/C/D/E/F |  | 0/3/1/3/5/2 | 3/2/3/6/1/6 | 6/4/3/1/4/3 | 0/1/3/1/1/0 | 1/0/2/0/0/0 |

*[1]Classes obtenues par analyse en composante principale suivie d'une classification ascendante hiérarchique. Comparaisons de la valeur de la classe par rapport à la moyenne de l'ensemble des individus (fonction catdes de FactoMineR). En gras, différences significatives ; * : P<0,05 ; ** : P<0,01 ; *** : P<0,001.*

### 2.4.3. Classification des truies selon leur comportement

La classification aboutit à cinq 'classes' de truies (Tableau 3).

**C1 :** Les 14 truies de cette classe sont jeunes (rang portée moyen de 2,2), changeant en moyenne 43 fois de posture par jour, soit près de 50% de plus que la valeur moyenne pour l'ensemble des truies. La durée des séquences moyennes et maximales couchées et debout sont toutes plus faibles que la moyenne des individus. En moyenne, la durée d'une séquence debout n'est que de 11 minutes, et 58 minutes pour la séquence couchée.

**C2 :** Les 21 animaux de cette classe passent peu de temps debout ; TpsD est inférieur de 1 h 26 par rapport à la valeur moyenne. Les séquences maximale et moyenne debout sont faibles.

**C3 :** Ces 21 truies présentent un temps debout élevé, supérieur de 68 minutes à la valeur moyenne, avec une séquence maximale debout de 2 h 20.

**C4 :** Les 6 truies de cette classe sont âgées (rang de portée de 6,5) et changent peu de posture. Les séquences couchées maximale et moyenne sont plus longues que la valeur moyenne de 53% et 54%, respectivement.

**C5 :** Cette classe rassemble 3 truies qui passent en moyenne 7 h 40 debout par jour, avec une séquence moyenne debout de 1 h 16 soit 54 minutes de plus que la valeur moyenne des truies. La séquence maximale debout est la plus élevée avec 5 h 21. Ces truies changent deux fois moins de posture que la valeur moyenne.

Des truies des différentes classes précédemment définies sont réparties dans les six élevages et ne dépendent ni de l'élevage, ni du type de conduite (groupe stable ou dynamique).

La figure 10 offre une autre représentation des données du tableau 3. Les cinq classes de truies sont positionnées sur un graphique à partir du temps moyen passé debout et du nombre de changements de posture.

**Figure 10 : Positionnement des 5 classes de truies selon le temps passé debout et le nombre de changements de posture par jour**

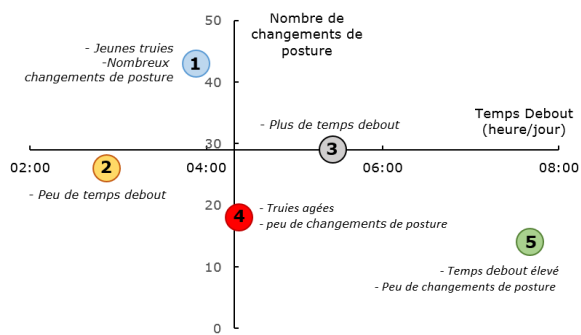

*Le graphique est centré sur les valeurs moyennes de l'ensemble des animaux (4h19 debout et 29 changements de posture)*

## 3. Discussion

### 3.1. Méthode de mesure utilisée

La fixation de capteurs à la patte de truies logées en groupes est un résultat très original qui a permis d'obtenir des données nouvelles sur l'activité des animaux. Le premier challenge sur ce projet consistait à fixer des capteurs sur des truies, et s'assurer de leur tenue dans le temps.

La fixation des accéléromètres sur la patte de la truie présente une difficulté majeure dans la mesure où les boîtiers métalliques peuvent être mordillés par les autres truies du groupe. Notre matériel était également volumineux, ce qui facilite sa préhension par l'animal.

La technique retenue à l'aide de collier de serrage en plastique correspond à un bon compromis entre la solidité du serrage, la facilité de pose sur des animaux logés en groupes, et le prix réduit du matériel de fixation. Ce matériel est resté en place pour une durée allant jusqu'à 5 jours consécutifs, et aurait probablement pu être laissé sur une durée plus longue dans plusieurs élevages. L'essentiel des pertes de matériel étaient observées au cours des premières 24 heures de mesures. Dans au moins un des élevages, le niveau de serrage insuffisant des colliers en plastique explique en partie la perte des boîtiers.

A notre connaissance, les accéléromètres ont été testés principalement sur la patte de truies logées en stalles individuelles (Ringgenberg et al., 2010), placés à l'oreille de la truie (Marchioro et al., 2011) ou sur des colliers autour du cou de truies (Cornou et Lundbye-Christensen, 2008). L'identification de comportements tels que l'alimentation, la fouille, la marche ou la distinction d'un couché ventral ou latéral à partir d'accéléromètres placés au niveau du cou demande l'utilisation de modèles mathématiques complexes (Cornou et Lundbye-Christensen, 2008 ; Escalante et al., 2013).

L'avantage principal de la fixation à la patte est la qualité de l'enregistrement et la facilité du traitement du signal. L'utilisation du seul axe vertical de l'accéléromètre apparaît suffisante pour mesurer les positions debout et couchée. Si une donnée plus précise était indispensable pour quantifier et qualifier le comportement de l'animal, par exemple par rapport à l'évaluation du besoin alimentaire, les deux autres axes de l'appareil pourraient être utilisés dans cet objectif.

### 3.2. Niveau d'activité des truies

Les mesures en élevage de production soulignent l'importance du temps passé debout par les truies et surtout l'importante variabilité entre élevages, et entre truies au sein du groupe.

Le temps moyen passé debout par les truies de notre étude est de 4h19, soit une valeur proche des 4h retenues en référence dans InraPorc pour évaluer le besoin alimentaire des animaux (Dourmad et al., 2005). Dans une optique d'adaptation de la distribution alimentaire aux besoins de l'animal, la connaissance de l'activité individuelle semble nécessaire compte tenu de l'importante variabilité entre les truies.

Le nombre de changements de posture apparaît comme un critère intéressant pour qualifier le comportement de l'animal. Avec 29 changements de posture enregistrés en moyenne par période de 24 heures, nos résultats sont du même ordre de grandeur que celui rapporté par Marchant et Broom (1996). Au sein des groupes, certaines truies changent fréquemment de posture. C'est notamment le cas de jeunes truies de la classe C1 qui changent de posture en moyenne toutes les 30 minutes. Il pourrait être intéressant de vérifier, par des observations comportementales, si ces changements de posture sont des comportements spontanés des truies, ou le résultat d'interactions avec d'autres truies du groupe (Arey et Edwards, 1998). Le mélange de cochettes avec des truies plus âgées est une pratique par ailleurs identifiée comme à risque sur le plan des performances de reproduction (Boulot et al., 2011). La répétabilité du comportement de la truie au cours de plusieurs jours successifs ouvre également la possibilité d'utiliser les capteurs pour la détection automatisée des troubles de reproduction et des problèmes d'aplombs. Un changement du rythme d'activité peut signaler précocement l'émergence d'un problème de santé.

Dans notre étude, les six élevages sélectionnés avaient une configuration proche, avec des DAC installés sur caillebotis. Les différences en termes de comportement apparaissent également moins contrastées que ce que nous avions observé en évaluant les distances parcourues par les truies en groupe dans des logements très divers (Tertre et Ramonet, 2014). Néanmoins, nous observons un effet de l'élevage sur le temps passé debout par les truies. Cet effet est certainement lié en partie à la configuration du bâtiment et à la conduite du troupeau. Mais d'autres paramètres tel que la nature et la fréquence des contacts entre l'éleveur et son troupeau (Gonyou et al., 1986) seraient également à considérer. La réaction des truies à notre approche pour la pose des capteurs était ainsi très différente dans les six élevages.

## 4. Conséquences pratiques en élevage

Le besoin énergétique d'entretien est doublé lorsque la truie gestante est debout, comparativement à la posture couchée (Noblet et al., 1994). Le niveau d'activité physique est une des causes proposée pour expliquer le syndrome de la truie maigre. Il s'agit souvent au départ d'un amaigrissement excessif de lactation non compensé en gestation et qui s'aggrave avec le temps et avec les parités (Martineau et Klopfenstein, 1996). Cariolet et Dantzer (1984) ont observé, sur des truies attachées, que les plus maigres étaient aussi les plus actives. Sur la base de ces résultats, Noblet et al. (1994) suggèrent que la truie est alors placée dans un cercle vicieux. L'activité physique de la truie maigre entraine une dépense énergétique supplémentaire qui l'empêchera de reconstituer ses réserves corporelles.

La conséquence est qu'il peut être nécessaire de revoir les rations des truies en fonction du niveau de leur activité physique. Actuellement, compte tenu du peu de références disponibles sur le temps passé debout et couché pour des truies en groupe, il est compliqué pour l'éleveur et son conseiller en alimentation d'ajuster les plans d'alimentation des animaux en fonction de ce facteur. Pour une truie de 250 kg, 2 heures d'activités supplémentaires par jour correspondent à un besoin de 165 g d'aliment (tableau 4), soit plus de 15 kg d'aliment sur l'ensemble de la gestation (Quiniou et al., 2014).

**Tableau 4 : Besoin alimentaire (en équivalent g/jour[1]) nécessaire pour couvrir les besoins liés à l'activité physique**

|  |  | Poids de la truie (kg) | | |
|---|---|---|---|---|
|  |  | 150 | 200 | 250 |
| Activité supplémentaire, par jour[2] | + 1 h | 60 | 70 | 85 |
|  | + 2 h | 115 | 140 | 165 |
|  | + 3 h | 170 | 210 | 250 |

[1] *Exprimé en g d'aliment formulé à 9,0 MJ EN/kg (d'après Quiniou et al., 2014).*

[2] *En supposant qu'une truie calme reste 2 heures debout en moyenne.*

Des simulations ont été réalisées avec le logiciel InraPorc pour des truies qui reçoivent une même quantité d'aliment pour des niveaux d'activité physique différenciés. Le figure 5 illustre le différentiel de prise de poids, mais plus encore les différences en termes d'évolution de l'épaisseur de lard dorsal au cours de 5 gestations successives. Sur plusieurs cycles de gestation, l'épaisseur de lard de la truie active (debout 360 minutes/j) est en diminution pour atteindre 13 mm à l'entrée en maternité du 5ème cycle. Pour des truies moyennement active (240 minutes/jour, référence de base InraPorc), l'épaisseur de lard se maintient à 20-22 mm, alors que pour la truie peu active (debout 180 min/jour) l'ELD continue d'augmenter. Cette simulation illustre le syndrome de la truie maigre hyperactive, incapable de reconstituer ses réserves corporelles au cours des cycles successifs.

Dans notre étude, le chronométrage du temps passé debout par les truies montre que des différences importantes existent entre les élevages, et entre les animaux au sein du même élevage. Les valeurs extrêmes sur le temps passé debout vont bien au-delà des données retenues dans les simulations précédentes. Pour l'éleveur qui a modifié il y a peu de temps les caractéristiques de ses salles de gestation, cette mesure n'est pas réalisable pratiquement mais le suivi de l'évolution de l'état de ses truies au cours des gestations, notamment vers un amaigrissement alors que les rations allouées sont restées stables, peut être un indicateur de nécessité à corriger les rations alimentaires. Certains éleveurs de notre étude réalisent jusqu'à 4 mesures d'épaisseur de lard au cours de chaque cycle de reproduction pour vérifier si l'évolution de l'état de la truie correspond aux attendus.

*Accéléromètres pour la mesure de l'activité physique des truies en groupes*

**Tableau 5 : Simulation de l'évolution du poids de la truie et de son épaisseur de lard dorsal sur 3 gestations successives, selon le niveau d'activité physique (d'après Quiniou, données non publiées)**.

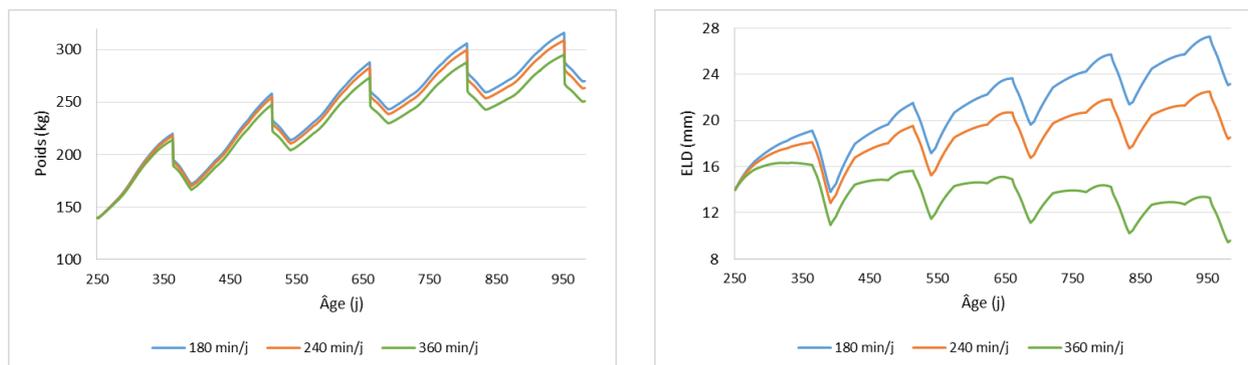

*Simulations réalisées sur InraPorc. Plan d'alimentation en U, différencié pour les cochettes, primipares et multipares. Aliments formulés à 9,5 MJ EN (cochette), 9,2 MJ EN(multipares), 9,8 MJ EN(Allaitante).*

A moyen terme, des équipements pourraient être disponibles pour mesurer en temps réel l'activité des animaux et modifier les rations en conséquence. Notre projet, à l'aide d'accéléromètres fixés à la patte, constitue une première étape dans cette direction. Un projet plus conséquent qui implique l'Ifip, l'Inra, la CRAB et deux équipementiers ambitionne de développer des outils de mesure de l'activité des truies en élevage et de correction des rations, pour amener vers une alimentation de précision pour les truies gestantes.

## 5. Conclusion

Notre étude nous a permis de mesurer le temps passé debout et couché par des truies logées en groupe dans des bâtiments avec DAC au cours de plusieurs journées successives. La fixation à la patte de l'animal permet d'obtenir un signal net et facile à interpréter. Les méthodes de fixation du capteur et de traitement du signal associé sont opérationnelles pour des mesures dans nos conditions d'étude. En revanche, une utilisation commerciale à grande échelle de ce type de capteur demanderait de travailler sur la miniaturisation du matériel et la transmission en continu des données.

Ce travail complète nos résultats obtenus sur l'activité et la distance parcourue par les truies logées en groupes, obtenus sur un temps court et nécessairement limité lors d'observations directes (Tertre et Ramonet, 2014). L'utilisation de capteurs fixés sur l'animal permet d'augmenter le temps de mesure et de diversifier les critères d'analyse au-delà de la seule posture. L'interprétation des données issues des capteurs repose sur des observations comportementales nécessaires pour qualifier en paramètre d'importance zootechnique le changement du rythme ou du profil de l'activité des animaux.

## 6. Pour plus d'informations

Contact : Yannick Ramonet

Pôle porc-aviculture des Chambres d'agriculture de Bretagne, Plérin

Tél. : 02 96 79 21 90

Mail : yannick.ramonet@bretagne.chambagri.fr

## 7. Références bibliographiques

*Accéléromètres pour la mesure de l'activité physique des truies en groupes*

**Abstract**

**Use of accelerometers to measure physical activity of group-housed pregnant sows. Method development and use in six pig herds.**

The development of precision livestock farming which adjusts the food needs of each animal requires detailed knowledge of its behavior and particularly physical activity. Individual differences between animals can be observed for group-housed sows. Accelerometer technology offers opportunities for automatic monitoring of animal behavior. The aim of the first step was to develop a methodology to attach the accelerometer on the sow's leg, and an algorithm to automatically detect standing and lying posture. Accelerometers (Hobo Pendant G) were put in a metal case and fastened with two cable ties on the leg of 6 group-housed sows. The data loggers recorded the acceleration on one axis every 20 s. Data were then validated by 9 hours of direct observations. The automatic recording device showed data of high sensitivity (98.8%) and specificity (99.8%). Then, accelerometers were placed on 12 to 13 group-housed sows for 2 to 4 consecutive days in 6 commercial farms equipped with electronic sow feeding. On average each day, sows spent 259 minutes (± 114) standing and changed posture 29 (± 12) times. The sow's standing time was repeatable day to day. Differences between sows and herds were significant. Based on behavioral data, 5 categories of sows were identified. This study suggests that the consideration of individual behavior of each animal would improve herd management.

**Comment citer ce document ?**

Yannick Ramonet, Carole Bertin, 2015. Utilisation d'accéléromètres pour évaluer l'activité physique des truies gestantes logées en groupes. Développement de la méthode et utilisation dans six élevages au DAC. Rapport d'étude. Chambres d'agriculture de Bretagne, 12 pages.

**Ce travail a donné lieu à 2 publications**

*Accéléromètres pour la mesure de l'activité physique des truies en groupes*

- Bertin C., Ramonet Y., 2015. Utilisation d'accéléromètres pour mesurer l'activité physique des truies logées en groupes. Développement de la méthode et utilisation dans six élevages. Journées de la Recherche Porcine, 47 : 229-234.
- Ramonet Y., Bertin C., 2015. Un capteur à la patte pour mesurer l'activité physique des truies. TechPorc, 21 : 19-22.

**Mots-clés : truie gestante, comportement, activité motrice**